# Effect of different substrates on Compact stacked square Microstrip Antenna

Asok De, N.S. Raghava, Sagar Malhotra, Pushkar Arora, Rishik Bazaz

**Abstract**— Selection of the most suitable substrate for a Microstrip antenna is a matter of prime importance. This is because many limitations of the microstrip antenna such as high return loss, low gain and low efficiency can be overcome by selecting an appropriate substrate for fabrication of the antenna, without shifting the resonant frequency significantly. The substate properties such as its dielectric constant, loss tangent have a pronounced effect on the antenna characteristics. Some of the critical properties that are to be taken care of while selecting a dielectric are homogeneity, moisture absorption and adhesion of metal- foil cladding. In this paper a comprehensive study of the effect of variation of substrate material on the antenna properties has been presented.

———————————— ◆ ————————————

## 1 INTRODUCTION

Substrate materials play an important role in antenna design, production and finished product performance.

A simple method that can be employed to modify the different properties of the antenna is by changing the substrate; as height and dielectric constant of the substrate influence the antenna properties.

The substrate in microstrip antenna is primarily required for giving mechanical strength to antenna [1]. The dielectric used is also responsible for degraded electrical properties of antenna as the surface waves produced on the dielectric extract a part of total power available for direct radiation (space waves) [2]. The cost incurred in the production of microstrip antenna is closely affected by the substrate used in its design. Therefore an intelligent decision has to be taken while selecting a substrate so as to satisfy both electrical and mechanical requirement for the antenna. The substrate properties that are taken into consideration while selecting a dielectric include: dielectric constant and loss tangent and their variation with temperature and frequency, homogeneity, dimensional stability with processing and temperature, humidity and aging. Other physical properties such as resistance to chemicals, impact resistance, formability, bonding ability, Foil adhesion (force needed to peel an etched strip of clad foil perpendicularly from an etched surface) etc, are important in fabrication [3].

In this paper study has been conducted on an EBG backed CSSMA with seven different commonly available dielectrics as its substrate. The proposed antenna can be used for WLAN applications and Road Vehicle Communications [4], [5].

• *Asok De is with the DelhiTechnological University, Delhi, India.*
• *N. S. Raghava is with the the DelhiTechnological University, Delhi, India.*
• *Sagar Malhotra is with the DelhiTechnological University, Delhi,India*
• *Pushkar Arora is with the DelhiTechnological University, Delhi,India*
• *Rishik Bazaz is with the DelhiTechnological University, Delhi,India*

## 2 ANTENNA DESIGN

A Compact Stacked Square Microstrip Antenna (CSSMA) backed with Electromagnetic band gap (EBG) structure is designed [4], [5]. The antenna consists of two layers of dielectric, the air dielectric is used as first layer and the dielectric material of the second or lower layer is kept variable. Two square patches of dimension 16mm x 16mm have been mounted over the substrate to form a stacked structure. The thickness (H1) of the upper dielectric (air) is 2.6mm and that of the lower one (H2) is 1.6mm.The top view of CSSMA is shown Fig.1

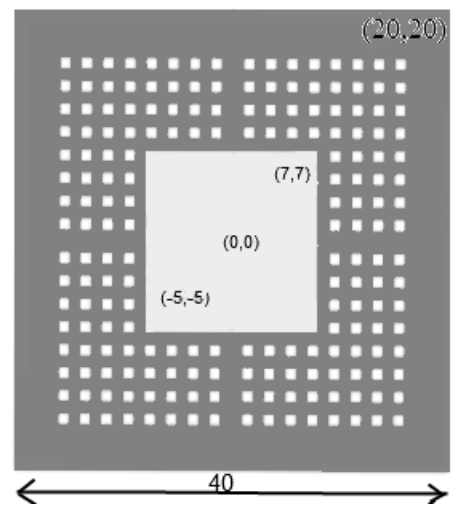

Fig. 1. Top view of CSSMA

The antenna is excited by a coaxial feed of diameter 1.3mm given on the lower patch at    (5mm x 5mm). A shorting post of diameter 1mm at (7mm x 7mm) is placed between the two patches.

To construct an EBG structure, a metallic square patch of side 40mm is constructed at the bottom side of the lower substrate. This metallic square serves as the ground plane. Square slots of side 1mm x 1mm spaced by 2mm have been etched on the ground plane at a distance of 4mm from edges with the constraint that there should be no holes beneath the patch.

A reflector plane is placed at a distance of 8.5mm behind the ground plane. The three dimensional view of the structure is shown in the Fig.2. The main advantage of using EBG structure is elimination of surface wave currents which are responsible for low antenna efficiency and degraded pattern. The reflector plane provided at the back helps in reducing the back lobe.

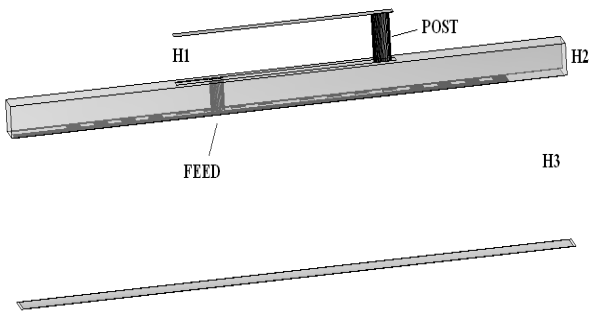

Fig. 2. 3D View of CSSMA

## 4. RESULT AND DISCUSSIONS

Study has been conducted to observe the changes in different parameters of CSSMA with the usage of seven different dielectrics commonly available in the market. The results of simulations are tabulated in Table 1.

TABLE 1
SIMULATION RESULTS

| Material | Electrical Permittivity (εr) | Loss Tangent | Resonant frequency (GHz) | Return Loss(S11) | Antenna Efficiency (%) | Gain (dBi) | Direc-tivity (dBi) |
|---|---|---|---|---|---|---|---|
| Foam | 1.05 | 0 | 2.707 | -4.145 | 61 | 2.99 | 5.09 |
| Duroid | 2.2 | 0.0009 | 2.64 | -11.88 | 93.51 | 4.47 | 4.76 |
| Benzo-cyclo-buten | 2.6 | 0 | 2.636 | -14.01 | 96.51 | 4.63 | 4.78 |
| Roger 4350 | 3.48 | 0.004 | 2.586 | -25.29 | 99.66 | 4.62 | 4.63 |
| Epoxy | 3.6 | 0 | 2.576 | -17.37 | 98.34 | 4.54 | 4.61 |
| FR4 | 4.4 | 0.018 | 2.556 | -24.08 | 99.6 | 4.54 | 4.56 |
| Glass | 5.5 | 0 | 2.535 | -19.14 | 98.88 | 4.44 | 4.49 |
| Duroid 6010 | 10.2 | 0.0023 | 2.455 | -9.449 | 88.64 | 4.02 | 4.55 |

It is observed that the resonant frequency slightly decreases with increase in electrical permittivity of the substrates used. Directivity reached its peak value (5.09) with Foam as substrate, but taking into consideration the corresponding values of return loss, antenna efficiency, and gain, antenna having foam dielectric gave the most unsatisfactory results amongst all dielectrics used. Despite the fact that the values of gain and directivity for Duroid and Duroid 6010 are well within the acceptable limits, their use is limited owing to their high return loss low level of efficiency. For the Roger's 4350 dielectric with 3.48 as electric permittivity both antenna efficiency and return loss reached to their maximum value. Although gain and directivity of benzocyclobuten dielectric are slightly higher than Roger's 4350 dielectric but considering the high value of efficiency and return loss of the latter, it is recommended to use Roger's 4350 as substrate. Glass, epoxy, and FR4 though giving slightly inferior results than Roger's 4350 has the advantage of being much more cost effective than the latter.

## 4. CONCLUSION

Various parameters of CSSMA have been studied by using different substrates. It is found that by selecting a suitable substrate specific antenna requirements can be met. It can be concluded that Roger's 4350 is the most efficient amongst the seven dielectrics used in CSSMA and has satisfactory value of gain and directivity for use in WLAN and Road Vehicle Communications.

**Asok De** received PhD from IIT Kharagpur in the area of Microstrip antenna in the year 1984. At present he is Professor in Electronics and Communication Engineering, Delhi Technological University (formerly Delhi College of Engineering). Presently he is Principal, Ambedkar Institute of Technology, Delhi. His field of interest is antennas, Numerical techniques in Electromagnetic, Transmission lines etc. He has published many research papers in reputed Journals.

**N. S. Raghava** is presently working as Lecturer (Selection Grade) in the Department of Electronics and Communication Engineering (ECE) of Delhi Technological University (formerly Delhi College of Engineering), Delhi. His teaching and research interests are in the areas of Analog and Digital Communications, Microstrip antennas and Digital Signal Processing.

**Pushkar Arora** is currently pursuing B.E. in Electronics and Communications Engineering in Delhi Technological University, Delhi, India.

**Sagar Malhotra** is currently pursuing B.E. in Electronics and Communications Engineering in Delhi Technological University, Delhi, India.

**Rishik Bazaz** is currently pursuing B.E. in Electronics and Communications Engineering in Delhi Technological University, Delhi, India.